\documentclass[aps,pra,twocolumn,showpacs,floatfix,superscriptaddress]{revtex4-1}

  \usepackage[utf8]{inputenc}
  \usepackage[T1]{fontenc}
  \usepackage{float} 
  \usepackage{color}  

\usepackage{graphicx}
\usepackage{amsmath}
\usepackage{amssymb}
\usepackage{mathrsfs}
\usepackage{bbm}
\usepackage{indentfirst}
\usepackage{dcolumn}
\usepackage{soul}

\begin{document}

\title{Formation and stability of metastable skyrmionic spin structures with various topologies in an ultrathin film}

\author{Levente R\'{o}zsa}
\email{rozsa.levente@wigner.mta.hu}
\affiliation{Institute for Solid State Physics and Optics, Wigner Research Centre for Physics, Hungarian Academy of Sciences,
P.O. Box 49, H-1525 Budapest, Hungary}
\author{Kriszti\'{a}n Palot\'{a}s}
\affiliation{Department of Theoretical Physics, Budapest University of Technology and Economics, Budafoki \'{u}t 8, H-1111 Budapest, Hungary}
\affiliation{Center for Computational Materials Science, Department of Complex Physical Systems, Institute of Physics, Slovak Academy of Sciences, SK-84511 Bratislava, Slovakia}
\author{Andr\'{a}s De\'{a}k}
\affiliation{Department of Theoretical Physics, Budapest University of Technology and Economics, Budafoki \'{u}t 8, H-1111 Budapest, Hungary}
\affiliation{MTA-BME Condensed Matter Research Group, Budapest University of Technology and Economics, Budafoki \'{u}t 8, H-1111 Budapest, Hungary}
\author{Eszter Simon}
\affiliation{Department of Theoretical Physics, Budapest University of Technology and Economics, Budafoki \'{u}t 8, H-1111 Budapest, Hungary}
\author{Rocio Yanes}
\affiliation{University of Salamanca, E-37008 Salamanca, Spain}
\author{L\'{a}szl\'{o} Udvardi}
\author{L\'{a}szl\'{o} Szunyogh}
\affiliation{Department of Theoretical Physics, Budapest University of Technology and Economics, Budafoki \'{u}t 8, H-1111 Budapest, Hungary}
\affiliation{MTA-BME Condensed Matter Research Group, Budapest University of Technology and Economics, Budafoki \'{u}t 8, H-1111 Budapest, Hungary}
\author{Ulrich Nowak}
\affiliation{Department of Physics, University of Konstanz, D-78457 Konstanz, Germany}
\date{\today}

\begin{abstract}

We observe metastable localized spin configurations with topological charges ranging from $Q=-3$ to $Q=2$ in a (Pt$_{0.95}$Ir$_{0.05}$)/Fe bilayer on Pd$(111)$ surface by performing spin dynamics simulations, using a classical Hamiltonian parametrized by \textit{ab initio} calculations. We demonstrate that the frustration of the isotropic exchange interactions is responsible for the creation of these various types of skyrmionic structures. The Dzyaloshinsky--Moriya interaction present due to the breaking of inversion symmetry at the surface energetically favors skyrmions with $Q=-1$, distorts the shape of the other objects, and defines a preferred orientation for them with respect to the underlying lattice.

\end{abstract}

\maketitle

\section{Introduction\label{sec1}}

Magnetic skyrmions correspond to localized spin configurations, where the directions of the magnetic moments span the whole unit sphere\cite{Nagaosa}. Due to their small size and the ability to set them into motion with significantly smaller current densities than magnetic domain walls\cite{Iwasaki,Jonietz}, they hold promising aspects as bits of information in future magnetic logic and memory devices\cite{Fert,Iwasaki2,Zhou}. Recently, their creation and manipulation was also demonstrated experimentally under room-temperature environments\cite{Jiang,Woo}.

Although localized spin configurations also exist as metastable states in the two-dimensional scale-free Heisenberg model\cite{Belavin}, stabilizing the radius of magnetic skyrmions requires a further interaction term in the Hamiltonian. The possible candidates for such an interaction identified so far include the Dzyaloshinsky--Moriya interaction\cite{Dzyaloshinsky,Moriya,Bogdanov2}, the frustration of Heisenberg-type exchange interactions\cite{Okubo}, and four-spin interactions\cite{Heinze}. The magnetostatic dipolar interaction is also capable of stabilizing circular magnetic bubble domains in thin films\cite{Malozemoff}, the size and shape of which can be more easily manipulated by the geometry of the system and external magnetic fields than in the case of Dzyaloshinsky--Moriya skyrmions\cite{Yu3,Yu4,Kiselev}.

Past investigations of skyrmions have mostly focused on Dzyaloshinsky--Moriya systems. This type of interaction only appears in noncentrosymmetric crystals, and is caused by the spin--orbit coupling. The skyrmion lattice phase was first identified in MnSi\cite{Muhlbauer}, and later in other bulk materials belonging to certain symmetry classes\cite{Yu2,Wilhelm,Munzer,Yu,Adams,Kezsmarki,Tokunaga}. While the skyrmion lattice is a thermodynamic phase\cite{Bogdanov,Dupe,Simon}, skyrmions may also appear as metastable localized spin configurations on the collinear background\cite{Leonov2}, and most suggested future applications rely on such individual or isolated skyrmions\cite{Fert,Iwasaki2,Zhou}. The presence of individual skyrmions has been demonstrated in several ultrathin and multilayer films\cite{Romming,Hsu,Moreau-Luchaire} by combining magnetic transition metals with heavy nonmagnetic elements, which provides a way of enhancing the Dzyaloshinsky--Moriya interaction\cite{Dupe2,Yang}.

Localized topological spin configurations may be classified according to the topological charge $Q$ and the helicity $\gamma$\cite{Nagaosa}. In Dzyaloshinsky--Moriya systems a given rotational sense of the spins is preferred, which selects a fixed value of topological charge and helicity for magnetic skyrmions\cite{Leonov2}. In contrast, dipolar systems allow for two helicity values which are degenerate in energy\cite{Yu3}, while in frustrated systems a continuous degeneracy arises\cite{Leonov,Lin}. Furthermore, the presence of biskyrmions (bound pairs of skyrmions) was recently demonstrated in several centrosymmetric materials\cite{Yu4,Lee,Wang}. Skyrmionic structures with different topological charges have also been identified in numerical calculations and simulations performed for frustrated systems\cite{Lin,Leonov,Dupe3}.

Besides the Dzyaloshinsky--Moriya interaction, the presence of frustrated isotropic exchange interactions has also been demonstrated recently in several ultrathin film systems\cite{Dupe3,Rozsa3,Romming3}. In particular, skyrmions have been observed in numerical simulations performed for (Pt$_{1-x}$Ir$_{x}$)/Fe bilayer on Pd$(111)$ in Ref.~\cite{Rozsa4}, and it was shown that the competition between ferromagnetic and antiferromagnetic isotropic exchange interactions is sufficiently strong to create an oscillating skyrmion--skyrmion interaction potential, previously only calculated for frustrated centrosymmetric systems\cite{Leonov,Lin}.

In this paper, we discuss the stability properties of metastable spin configurations with different topological charges found in the collinear field-polarized or ferromagnetic state of (Pt$_{0.95}$Ir$_{0.05}$)/Fe/Pd$(111)$. We perform spin dynamics simulations based on the Landau--Lifshitz--Gilbert equation, and by using a model Hamiltonian for the system parametrized by \textit{ab initio} calculations in Ref.~\cite{Rozsa4}. The Dzyaloshinsky--Moriya interaction present in the system selects skyrmions with $Q=-1$ as the energetically most favorable spin configuration. Here we will demonstrate that the frustrated exchange interactions are also capable of stabilizing localized spin configurations with topological charges $Q=-3,-2,0,1,$ and $2$, although the Dzyaloshinsky--Moriya interaction deforms their shape.

The paper is organized as follows. We summarize the theoretical background in Sec.~\ref{sec2}: in Sec.~\ref{sec2a} we present the parameters of the model Hamiltonian, and discuss the spin dynamics simulation method; while in Sec.~\ref{sec3} we introduce the topological charge $Q$, the vorticity $m$, and the helicity $\gamma$ in the continuum model, being the quantities that characterize the different types of skyrmionic structures. The results are presented in Sec.~\ref{sec4}: in Sec.~\ref{sec4a} we discuss the shape and the energy of the localized spin configurations; and in Sec.~\ref{sec5} we examine the preferred orientation of skyrmionic structures with respect to the lattice in detail. Finally, we summarize our results in Sec.~\ref{sec6}.


\section{Methods\label{sec2}}

\subsection{Spin model and spin dynamics\label{sec2a}}

For the description of the Fe magnetic moments in the (Pt$_{0.95}$Ir$_{0.05}$)/Fe/Pd$(111)$ ultrathin film, we have applied a model Hamiltonian with classical spins $\boldsymbol{S}_{i}$,
\begin{eqnarray}
H=\frac{1}{2}\sum_{i \ne j}\boldsymbol{S}_{i}\mathcal{J}_{ij} \boldsymbol{S}_{j}+\sum_{i}\boldsymbol{S}_{i}\mathcal{K} \boldsymbol{S}_{i}-\sum_{i}M\boldsymbol{S}_{i}\boldsymbol{B},\label{eqn1}
\end{eqnarray}
where $\boldsymbol{B}$ denotes the external magnetic field. The magnetic moment $M$, the coupling coefficients $\mathcal{J}_{ij}$ and the on-site anisotropy tensor $\mathcal{K}$ have been determined from \textit{ab initio} calculations based on the screened Korringa--Kohn--Rostoker method\cite{Szunyogh,Zeller} and the relativistic torque method\cite{Udvardi}. These calculations are reported in detail in Ref.~\cite{Rozsa4}.

In the following, lower case Greek letters will denote the Cartesian components of the spins. The isotropic coupling coefficients are defined as $J_{ij}=\frac{1}{3}\mathcal{J}_{ij}^{\alpha\alpha}$, leading to an energy expression of the form
\begin{eqnarray}
H_{\textrm{iso}}=\frac{1}{2}\sum_{i \ne j}J_{ij}\boldsymbol{S}_{i}\boldsymbol{S}_{j},\label{eqn2}
\end{eqnarray}
coinciding with the classical Heisenberg model. The antisymmetric part of the interaction tensor may be decomposed into a vector $D_{ij}^{\gamma}=\frac{1}{2}\varepsilon^{\alpha\beta\gamma}\mathcal{J}_{ij}^{\alpha\beta}$, with the energy expression
\begin{eqnarray}
H_{\textrm{DM}}=\frac{1}{2}\sum_{i \ne j}\boldsymbol{D} _{ij}\left(\boldsymbol{S}_{i}\times\boldsymbol{S}_{j}\right),\label{eqn3}
\end{eqnarray}
describing the Dzyaloshinsky--Moriya interaction. Finally, we mention that the difference between the diagonal components of $\mathcal{J}_{ij}$ will induce an energy difference between the out-of-plane and in-plane ferromagnetic orientations, which we will refer to as two-site anisotropy.

The interaction coefficients between the Fe spins are summarized in Table~\ref{table1}. The ground state of the system is a right-rotating cycloidal spin spiral state, which transforms into the collinear field-polarized state when a magnetic field of $B=0.21\,\textrm{T}$ is applied perpendicularly to the surface\cite{Rozsa4}. We will identify the topological objects in this field-polarized state. 

\begin{table}
  \centering
\begin{ruledtabular}
    \begin{tabular}{rrr}
    $d$ [$a$] & $J_{ij}$ [mRy] & $D^{\Vert}_{ij}$ [mRy] \\
    \hline
    1.0000 & -1.6952 & 0.0896 \\
    1.7321 & 0.1525 & -0.0037 \\
    2.0000 & 0.4250 & -0.0576 \\
    2.6458 & -0.0477 & -0.0114 \\
    3.0000 & -0.0453 & 0.0233 \\
    3.4641 & -0.0035 & -0.0025 \\
    3.6056 & 0.0285 & 0.0053 \\
    4.0000 & 0.0275 & -0.0041 \\
    4.3589 & 0.0014 & -0.0019 \\
    4.5826 & -0.0045 & 0.0015 \\
    5.0000 & -0.0169 & -0.0012 \\
    \end{tabular}%
\end{ruledtabular}
  \caption{Isotropic exchange interactions $J_{ij}$ and in-plane components of the Dzyaloshinsky--Moriya vectors $D^{\Vert}_{ij}$ between the Fe spins as a function of their distance $d$, given in terms of the lattice constant of the triangular lattice on the Pd$(111)$ surface ($a=2.751\,\textrm{\AA}$). $J_{ij}<0$ denotes ferromagnetic coupling, while $J_{ij}>0$ is antiferromagnetic. $D^{\Vert}_{ij}>0$ denotes that the Dzyaloshinsky--Moriya vector prefers the right-handed rotation of the spins, $D^{\Vert}_{ij}<0$ stands for left-handed rotation. The magnetic moment is $M=3.3\,\mu_{\textrm{B}}$. The total anisotropy energy between the out-of-plane and in-plane orientations is $\left(E_{\textrm{FM}}^{\bot}-E_{\textrm{FM}}^{\Vert}\right)/N=-0.0588\,\textrm{mRy}$, which includes both on-site and two-site contributions.\label{table1}}
\end{table}%

We have examined the possible spin configurations by numerically solving the Landau--Lifshitz--Gilbert equation\cite{Nowak},
\begin{eqnarray}
\frac{\textrm{d}\boldsymbol{S}_{i}}{\textrm{d}t}&=&-\gamma' \boldsymbol{S}_{i} \times \boldsymbol{B}_{i}^{\textrm{eff}} - \gamma' \alpha\boldsymbol{S}_{i} \times\left(\boldsymbol{S}_{i} \times \boldsymbol{B}_{i}^{\textrm{eff}}\right).\label{eqn4}
\end{eqnarray}
The parameters of the Hamiltonian (\ref{eqn1}) appear in the effective field $\boldsymbol{B}_{i}^{\textrm{eff}}=-\frac{1}{M}\frac{\partial H}{\partial \boldsymbol{S}_{i}}$. The dimensionless Gilbert damping coefficient is denoted by $\alpha$, while $\gamma'=\frac{\gamma}{1+\alpha^{2}}$ stands for the modified gyromagnetic ratio $\gamma=\frac{ge}{2m}$, with $g,e,m$ the electronic spin $g$ factor, absolute charge, and mass, respectively. The numerical integrations were performed by the semi-implicit B method from Ref.~\cite{Mentink}, which was primarily developed for finite-temperature calculations, but provides a sufficiently fast relaxation at zero temperature.

During the simulations it had to be ensured that the obtained localized spin configurations are indeed metastable, meaning that they represent a local energy minimum in configuration space and they cannot be destroyed by small rotations of the spins. On the other hand, since they possess a higher energy than the field-polarized ground state, they may get destroyed by spin fluctuations over long timescales, for example due to temperature effects. For this purpose, we initialized the system in a random state, then relaxed the spins by numerically solving Eq.~(\ref{eqn4}), which generally yielded a dilute array of different topological objects. Without thermal effects, this relaxation process corresponds to finding the nearest local energy minimum in configuration space, but not necessarily the ground state which is the global energy minimum. The speed of the relaxation is maximized by using $\alpha=1$. This method is similar to performing Monte Carlo simulations at almost zero temperature ($T=1\,\textrm{K}$) by starting from a random initial state as discussed in Ref.~\cite{Dupe3}, and corresponds to the infinitely fast limit of the rapid cooling process used in Ref.~\cite{Lin}.

We calculated the energy of the skyrmionic structures by cutting them out of the final configuration, positioning them on a field-polarized background of size $N=128\times128$ atoms with periodic boundary conditions, then performing another energy minimization simulation. As will be shown below, the characteristic size of the localized configurations was significantly smaller than the lattice size, so the effect of the boundary conditions was negligible. In all considered cases, we found that the decrease in energy was gradually slowing down over time, and we stopped the simulations when the energy of the system changed by less than $10^{-4}\,\textrm{mRy}$ over the last $12\,\textrm{ps}$. This procedure guaranteed that we indeed found a metastable state which the system cannot leave at zero temperature.

\subsection{Topological charge in the continuum model\label{sec3}}

We will introduce the quantities characterizing the localized spin configurations in the continuum description, following the notations of Refs.~\cite{Nagaosa,Leonov,Lin}. The Hamiltonian of the system in the micromagnetic model reads
\begin{eqnarray}
\mathscr{H}=&&\int \Big[-\mathscr{J}_{1}\left(\boldsymbol{\nabla}\boldsymbol{S}\right)^{2}+\mathscr{J}_{2}\left(\boldsymbol{\nabla}^{2}\boldsymbol{S}\right)^{2}+\mathscr{D}w_{\textrm{DM}}\left(\boldsymbol{S}\right)\nonumber
\\
&&-\mathscr{K}\left(S^{z}\right)^{2}-\mathscr{B}S^{z}\Big]\textrm{d}^{2}\boldsymbol{r},\label{eqn4a}
\end{eqnarray}
where the unit-length vector field $\boldsymbol{S}$ denotes the spins. The differentiation $\boldsymbol{\nabla}$ and the integral is understood in the two-dimensional plane $\boldsymbol{r}=\left(x,y\right)$, while the $z$ axis is identified with the out-of-plane direction. The first two terms with $\mathscr{J}_{1},\mathscr{J}_{2}>0$ describe the frustrated exchange interactions preferring a spin spiral with a finite wave vector\cite{Leonov,Lin,Michelson}. In the atomistic model, this frustration corresponds to the competition between the ferromagnetic nearest-neighbor interaction and the antiferromagnetic interaction with the second and third neighbors -- see Table~\ref{table1}, or Ref.~\cite{Rozsa4} for a more detailed discussion.

The third term in Eq.~(\ref{eqn4a}) stands for the Dzyaloshinsky--Moriya interaction, with
\begin{eqnarray}
w_{\textrm{DM}}\left(\boldsymbol{S}\right)=S^{z}\partial_{x}S^{x}-S^{x}\partial_{x}S^{z}+S^{z}\partial_{y}S^{y}-S^{y}\partial_{y}S^{z}\label{eqn5}
\end{eqnarray}
in the $C_{3\textrm{v}}$ symmetry class\cite{Bogdanov}. The fourth and fifth terms describe the presence of the out-of-plane easy axis ($\mathscr{K}>0$) and the external magnetic field $\mathscr{B}$.

For discussing the localized spin configurations, we will represent the spins in the spherical variables $\Theta$ and $\Phi$,
\begin{eqnarray}
\boldsymbol{S}=\left[\begin{array}{c}\sin\Theta\cos\Phi \\ \sin\Theta\sin\Phi \\ \cos\Theta\end{array}\right],\label{eqn6}
\end{eqnarray}
and the two-dimensional plane in polar coordinates $\boldsymbol{r}=\left(r,\varphi\right)$. The topological charge $Q$ is defined as\cite{Belavin}
\begin{eqnarray}
Q=\frac{1}{4\pi}\int\boldsymbol{S}\cdot\left(\partial_{x}\boldsymbol{S}\times\partial_{y}\boldsymbol{S}\right)\textrm{d}x\textrm{d}y,\label{eqn7}
\end{eqnarray}
which after performing the necessary change in the integration variables transforms into\cite{Nagaosa,Rozsa2}
\begin{eqnarray}
Q=\frac{1}{4\pi}\int_{0}^{\infty}\int_{0}^{2\pi}\left(\partial_{r}\Theta\partial_{\varphi}\Phi-\partial_{\varphi}\Theta\partial_{r}\Phi\right)\sin\Theta\textrm{d}\varphi\textrm{d}r.\label{eqn8}
\end{eqnarray}

Equation (\ref{eqn8}) counts how many times the vector field $\boldsymbol{S}$ winds around the unit sphere. We note that we have relied on a discretized version of Eq.~(\ref{eqn7}) during the spin dynamics simulations; see Refs.~\cite{Berg,Rozsa2} for details.

Following Refs.~\cite{Nagaosa,Lin}, in the next step we will assume that the $\Theta$ and $\Phi$ functions only depend on the variables $r,\varphi$ as $\Theta\left(r\right)$ and $\Phi\left(\varphi\right)$, the latter in the form
\begin{eqnarray}
\Phi\left(\varphi\right)=m\varphi+\gamma,\label{eqn9}
\end{eqnarray}
corresponding to circular spin configurations. The variable $m$ is called vorticity, while $\gamma$ is the helicity. In this case, Eq.~(\ref{eqn8}) may be expressed analytically as
\begin{eqnarray}
Q=-\frac{1}{2}\left[\cos\Theta\left(r\right)\right]_{0}^{\infty}\frac{1}{2\pi}\left[\Phi\left(\varphi\right)\right]_ {0}^{2\pi}=-m\:\textrm{sgn}\mathscr{B}.\label{eqn10}
\end{eqnarray}
For the magnetic field pointing outwards from the surface ($\textrm{sgn}\mathscr{B}=1$), the polar angle of the spins rotates from $\Theta=\pi$ in the origin to $\Theta=0$ in the field-polarized state, where the spins are parallel to the external field. This means that the localized configurations may be uniquely characterized by the vorticity $m$, which counts how many times and in which direction the in-plane components of the spins rotate around the circle when following a closed curve containing the origin on the surface. In the following, localized spin configurations with $m>0$ will be called skyrmions, in contrast to antiskyrmions with $m<0$\cite{Leonov}. Note that the topological charge $Q$ cannot be used for such a unique classification, because it also changes sign under time reversal.

\begin{figure*}
\centering
\includegraphics[width=2\columnwidth]{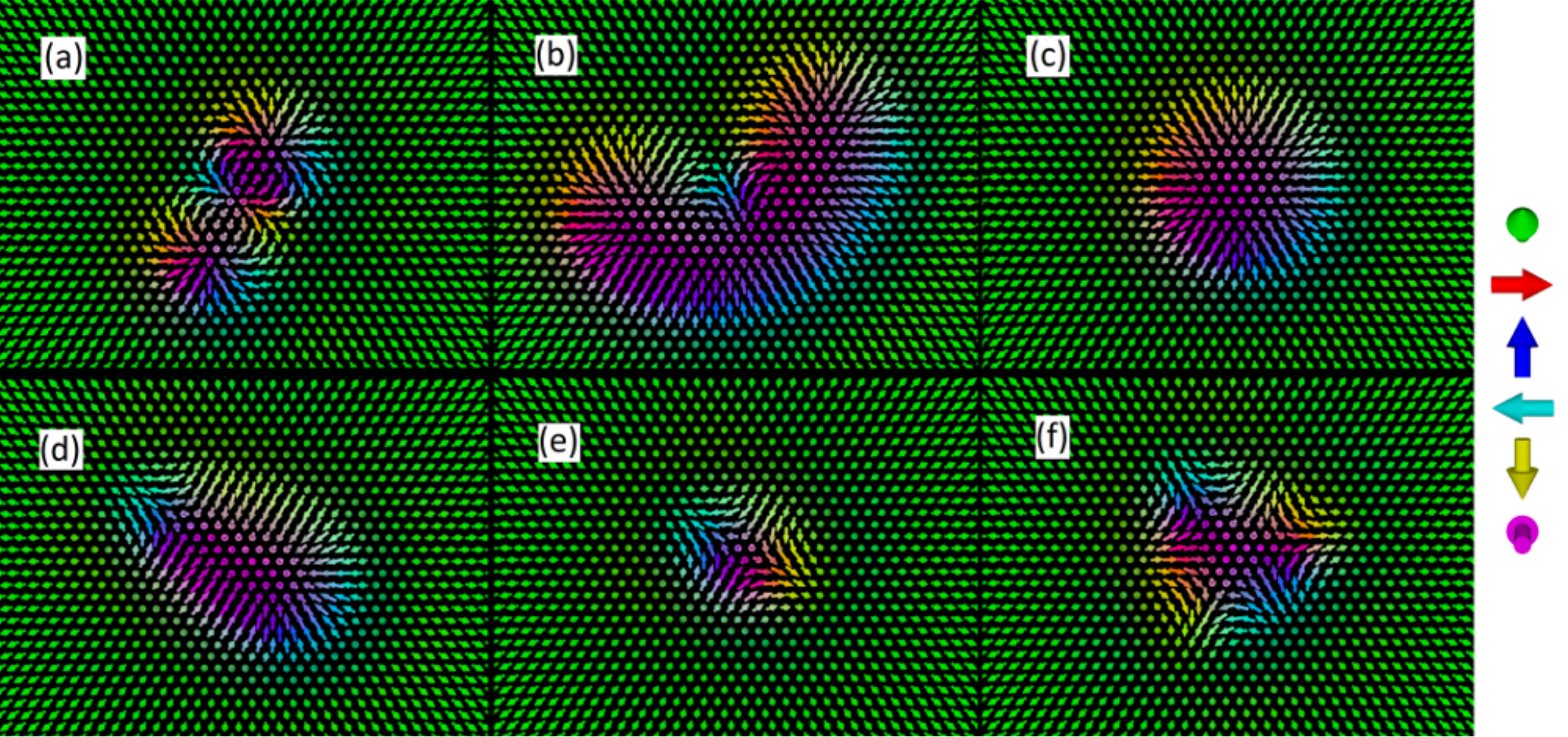}
\caption{Metastable localized spin configurations with different topological charges: (a) skyrmion with $Q=-3$, (b) skyrmion with $Q=-2$\cite{Yu4,Leonov,Lin}, (c) skyrmion with $Q=-1$, (d) ``chimera'' skyrmion with $Q=0$, (e) antiskyrmion with $Q=1$, and (f) antiskyrmion with $Q=2$\cite{Dupe3}. The value of the external field is $B=2.35\,\textrm{T}$ in part (a) and $B=0.23\,\textrm{T}$ in parts (b)-(f); the ground state is field-polarized for $B>0.21\,\textrm{T}$\cite{Rozsa4}. The colors indicate the directions of the spin vectors, illustrated on the right edge of the figure.\label{fig3}}
\end{figure*}

It was calculated in Ref.~\cite{Bogdanov} that the energy density of the Dzyaloshinsky--Moriya interaction in the transformed coordinates reads
\begin{eqnarray}
w_{\textrm{DM}}\left(\boldsymbol{S}\right)=&&\cos\left(\varphi-\Phi\right)\partial_{r}\Theta-\frac{1}{r}\sin\left(\varphi-\Phi\right)\partial_{\varphi}\Theta\nonumber
\\
&&+\sin\Theta\cos\Theta\sin\left(\varphi-\Phi\right)\partial_{r}\Phi\nonumber
\\
&&+\frac{1}{r}\sin\Theta\cos\Theta\cos\left(\varphi-\Phi\right)\partial_{\varphi}\Phi.\label{eqn11}
\end{eqnarray}

Equation~(\ref{eqn11}) indicates that in the presence of the Dzyaloshinsky--Moriya interaction, circular solutions as in Eq.~(\ref{eqn9}) may only be found for $m=1$; for other values of the vorticity the $r$ and $\varphi$ variables cannot be separated during the solution of the Euler--Lagrange equations constructed from Eq.~(\ref{eqn4a}). This means that the Dzyaloshinsky--Moriya interaction will distort the form of topological objects with $m\neq 1$.

On the other hand, it was demonstrated in Ref.~\cite{Lin} that the other terms in Eq.~(\ref{eqn4a}) admit circular solutions, skyrmionic structures with different values of $m$ may be stabilized, and the energy of the configuration will not depend on the sign of the vorticity. If the energy is calculated in such a circular configuration in the presence of the Dzyaloshinsky--Moriya interaction, it turns out that only skyrmions with $m=1$ gain energy from the chiral term due to the periodicity of the $\cos$ function in Eq.~(\ref{eqn11}).

It was calculated in Eq.~(\ref{eqn10}) that the topological charge does not depend on the helicity $\gamma$. Its role may be explained by rotating the spin configuration by the angle $\varphi_{0}$, which will transform $\Phi$ as
\begin{eqnarray}
\Phi'\left(\varphi\right)=\Phi\left(\varphi-\varphi_{0}\right)+\varphi_{0},\label{eqn12}
\end{eqnarray}
while leaving $\Theta$ unchanged. For $m=1$, this implies $\Phi'=\Phi$ for an arbitrary value of $\varphi_{0}$, meaning that the spin configuration is cylindrically symmetric, and that the helicity $\gamma$ is well-defined. The preferred value of the helicity minimizing the energy in the considered system is either $\gamma=0$ or $\gamma=\pi$, determined by the sign of $\mathscr{D}$ and the direction of the external field. Such a skyrmion is called a N\'{e}el skyrmion, in contrast to Bloch skyrmions with $\gamma\in\left\{\frac{\pi}{2},-\frac{\pi}{2}\right\}$\cite{Nagaosa}. This characterization refers to the type of spin rotation in the $360^{\circ}$ domain wall along an arbitrary cross-section going through the center of the skyrmion with $m=1$.

For all other values of the vorticity, rotating the configuration is equivalent to transforming the helicity as
\begin{eqnarray}
\gamma'=\gamma+\left(1-m\right)\varphi_{0}.\label{eqn13}
\end{eqnarray}

This means that for other localized configurations, the rotational sense of the spins is different along different cross-sections. This explains why they do not gain energy from the Dzyaloshinsky--Moriya interaction, since the latter selects a preferred rotational sense. From Eq.~(\ref{eqn13}) it can also be seen that skyrmionic structures with $m\neq 1$ possess a $C_{\left|1-m\right|}$ symmetry, in contrast to the cylindrical symmetry of the one with vorticity $m=1$.

\section{Results\label{sec4}}

\subsection{Shape and energy of localized spin configurations\label{sec4a}}

During the spin dynamics simulations we could identify six types of metastable localized spin configurations in the field-polarized state of (Pt$_{0.95}$Ir$_{0.05}$)Fe bilayer on Pd$(111)$, which are displayed in Fig.~\ref{fig3}, while their energies are summarized in Table~\ref{table2}. Note that the spins in the field-polarized state were oriented out-of-plane throughout the calculations, leading to the identification $Q=-m$.  We mention that some combinations of these skyrmionic structures could also be observed by using other sets of interaction parameters reported in Ref.~\cite{Rozsa4}, obtained for different concentrations of Ir in the overlayer. This indicates that the stabilization mechanism is connected to the general micromagnetic functional Eq.~(\ref{eqn4a}), not the precise values of the interaction parameters in Table~\ref{table1}.

In agreement with the considerations given in Sec.~\ref{sec3}, we found that the Dzyaloshinsky--Moriya interaction selects skyrmions with $Q=-1$ as the lowest-energy configuration. The energy gain is due to the fact that the rotational sense of the spins along any cross-section of skyrmions is right-handed, corresponding to the helicity value $\gamma=\pi$. All other topological objects become distorted compared to the circular approximation given in Eq.~(\ref{eqn9}) due to the Dzyaloshinsky--Moriya interaction. Although the rotational sense of the spins depends on the chosen cross-section, their distorted shape maximizes the energy gain from the energetically preferable right-handed rotation. This is clearly visible for the skyrmion with $Q=-2$, were the two constituent skyrmions with $Q=-1$ can be identified.

It can also be observed in Fig.~\ref{fig3} that the $C_{\left|1-m\right|}$ rotational symmetry of topological objects, which we have deduced in the circular approximation (see Eq.~(\ref{eqn13})), is conserved for the distorted skyrmionic structures. The skyrmion with $Q=-3$ and the antiskyrmion with $Q=1$ are both elongated, possessing a $C_{2}$ symmetry. The antiskyrmion with $Q=2$ has a mostly triangular shape, while the skyrmion with $Q=-2$ will only be transformed into itself after a rotation by $2\pi$. This also holds for the localized spin configuration with $Q=0$ in Fig.~\ref{fig3}(d). It consists of a ``head'' of a skyrmion and the ``tail'' of an antiskyrmion, and therefore we have named it a ``chimera'' skyrmion. Although it represents a metastable state, the ``chimera'' skyrmion is topologically equivalent to the field-polarized state, and consequently it is easy to collapse it by applying a higher value of the external magnetic field; for example, it is no longer stable at $B=2.35\,\textrm{T}$ given in Table~\ref{table2}. The in-plane magnetization component of the half-skyrmion and half-antiskyrmion points in the same direction, leading to a net in-plane magnetization for the ``chimera'' skyrmion. We note that the ``chimera'' skyrmion is similar to the topologically trivial magnetic bubble reported in Ref.~\cite{Yu4}, although significantly smaller in size.

\begin{table}
    \centering
\begin{ruledtabular}
    \begin{tabular}{rrrrrr}
    \multicolumn{3}{c}{$B=0.23\,\textrm{T}$} & \multicolumn{3}{c}{$B=2.35\,\textrm{T}$} \\
    \hline
    $Q$     & $E \left[\textrm{mRy}\right]$     & $E_{\textrm{b}} \left[\textrm{mRy}\right]$    & $Q$     & $E \left[\textrm{mRy}\right]$     & $E_{\textrm{b}} \left[\textrm{mRy}\right]$ \\
    \hline
    -1    & 0.82 &   n.a.    & -1    & 4.12 & n.a. \\
    -2    & 4.11 & 2.46 & -2    & 10.51 & 2.26 \\
    0     & 3.08 &   n.a.    & -3    & 15.16 & 2.79 \\
    1     & 5.11 &   n.a.    & 1     & 6.83 & n.a. \\
    2     & 7.92 & -2.30 & 2     & 11.51 & -2.16 \\
    \end{tabular}%
\end{ruledtabular}
\caption{Energy with respect to the field-polarized state $E$ of the spin configurations in Fig.~\ref{fig3}. The binding energy is calculated as $E_{\textrm{b}}\left(Q\right)=E\left(Q\right)-\left|Q\right|E\left(\textrm{sgn}Q\right)$, that is, by assuming that higher-order skyrmions and antiskyrmions represent bound states of $Q=\pm1$ units.\label{table2}}
\end{table}

\begin{figure*}[tbp]
\centering
\includegraphics[width=2\columnwidth]{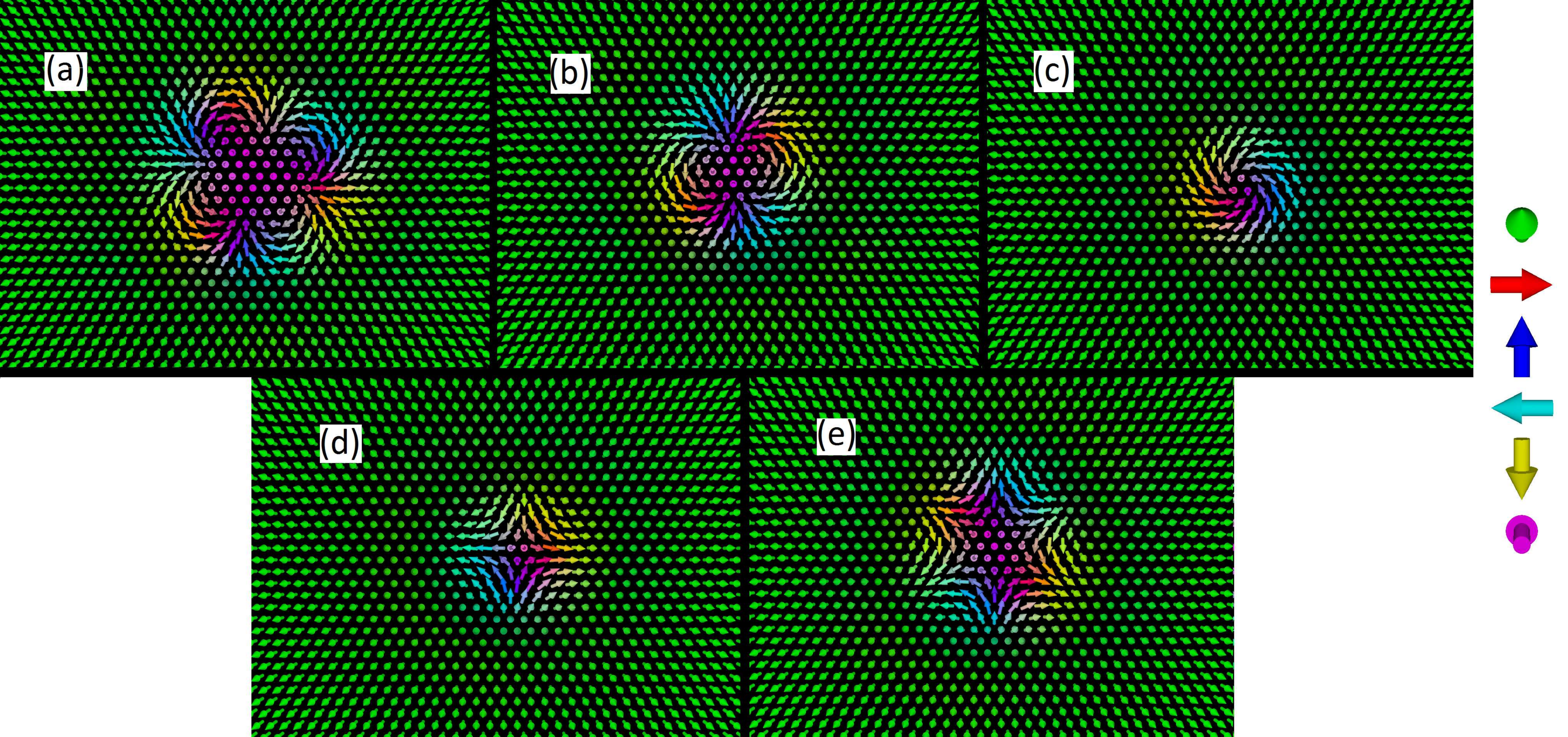}
\caption{Metastable localized spin configurations with topological charges (a) $Q=-3,$ (b) $Q=-2,$ (c) $Q=-1,$ (d) $Q=1$, and (e) $Q=2$. Compared to Fig.~\ref{fig3}, only the isotropic part of the exchange tensors was kept. The value of the external field is $B=0.23\,\textrm{T}$.\label{fig6}}
\end{figure*}

Skyrmions with $Q=-3,-2$ in Table~\ref{table2} have a positive binding energy; due to this reason, they can easily split into constituents with $Q=-1$. Higher-order skyrmions represent a lower magnetization difference with respect to the field-polarized state than two or three individual skyrmions; therefore, the positive binding energy slightly decreases as the external field is increased due to the energy gain from the Zeeman term -- see the row for $Q=-2$ in Table~\ref{table2}. Increasing the effect of the Zeeman term is necessary for stabilizing skyrmions with $Q=-3$; this is why a significantly higher value of the external magnetic field was used in Fig.~\ref{fig3}(a) than for the other configurations. On the other hand, the antiskyrmion with $Q=2$ possesses a negative binding energy, and consequently cannot split into two antiskyrmions with $Q=1$. This makes higher-order antiskyrmions more stable against increasing or decreasing the value of the external field compared to higher-order skyrmions.

The stability of localized spin configurations against thermal fluctuations is mainly determined by the energy barrier separating them from the field-polarized state instead of the relative and binding energies listed in Table~\ref{table2}. This energy barrier strongly depends on the magnetic field and system parameters\cite{Hagemeister,Lobanov}. In order to examine the relative stability of the skyrmionic structures, we performed finite-temperature spin dynamics simulations -- for the method see e.g. Ref.~\cite{Rozsa4}. We initialized the system in the relaxed metastable states found at zero temperature, and run the simulations for $484\,\textrm{ps}$ at selected temperature values between $T=4.7\,\textrm{K}$ and $T=15.8\,\textrm{K}$.

\begin{table}[tbp]
    \centering
\begin{ruledtabular}
    \begin{tabular}{rrrrrr}
    \multicolumn{3}{c}{$B=0.23\,\textrm{T}$} & \multicolumn{3}{c}{$B=2.35\,\textrm{T}$} \\
    \hline
    $Q$     & $E \left[\textrm{mRy}\right]$     & $E_{\textrm{b}} \left[\textrm{mRy}\right]$    & $Q$     & $E \left[\textrm{mRy}\right]$     & $E_{\textrm{b}} \left[\textrm{mRy}\right]$ \\
    \hline
    -1    & 5.41 &   n.a.    & -1    & 6.96 & n.a. \\
    -2    & 8.70 & -2.12 & -2    & 11.94 & -1.99 \\
    -3     & 12.54 &   -3.69    & -3    & 18.00 & -2.89 \\
    1     & 5.41 &   n.a.    & 1     & 6.96 & n.a. \\
    2     & 8.70 & -2.12 & 2     & 11.94 & -1.99 \\
    \end{tabular}%
\end{ruledtabular}
\caption{Energy with respect to the field-polarized state $E$ of the spin configurations in Fig.~\ref{fig6}. The binding energy is calculated as $E_{\textrm{b}}\left(Q\right)=E\left(Q\right)-\left|Q\right|E\left(\textrm{sgn}Q\right)$, that is, by assuming that higher-order skyrmions and antiskyrmions represent bound states of $Q=\pm1$ units.\label{table3}}
\end{table}

The net topological charge did not change during any of the simulations. The skyrmionic structures with $Q=-1,0,1,$ and $2$ remained stable with the interaction parameters and magnetic field values denoted in Fig.~\ref{fig3}. At higher fields, we found that the ``chimera'' skyrmion may collapse due to thermal fluctuations even if it was metastable at zero temperature. This is in agreement with the above argument; namely, that it is more sensitive to the value of the magnetic field than the other configurations. Skyrmions with $Q=-3,-2$ separated into individual skyrmions already at $T=4.7\,\textrm{K}$, probably because the height of the energy barrier is small due to the large positive binding energies of these structures. However, we found that these structures remained stable against thermal fluctuations at $B=0\,\textrm{T}$  in Pt/Fe/Pd$(111)$ instead of  (Pt$_{0.95}$Ir$_{0.05}$)/Fe/Pd$(111)$. Although the interaction parameters do not differ considerably between these two systems (see Ref.~\cite{Rozsa4} for a comparison), the binding energy of skyrmions with $Q=-3,-2$ is below $1\,\textrm{mRy}$ in Pt/Fe/Pd$(111)$ at $B=0\,\textrm{T}$, which is significantly lower than the values listed in Table~\ref{table2}.

\begin{figure*}
\centering
\includegraphics[width=2\columnwidth]{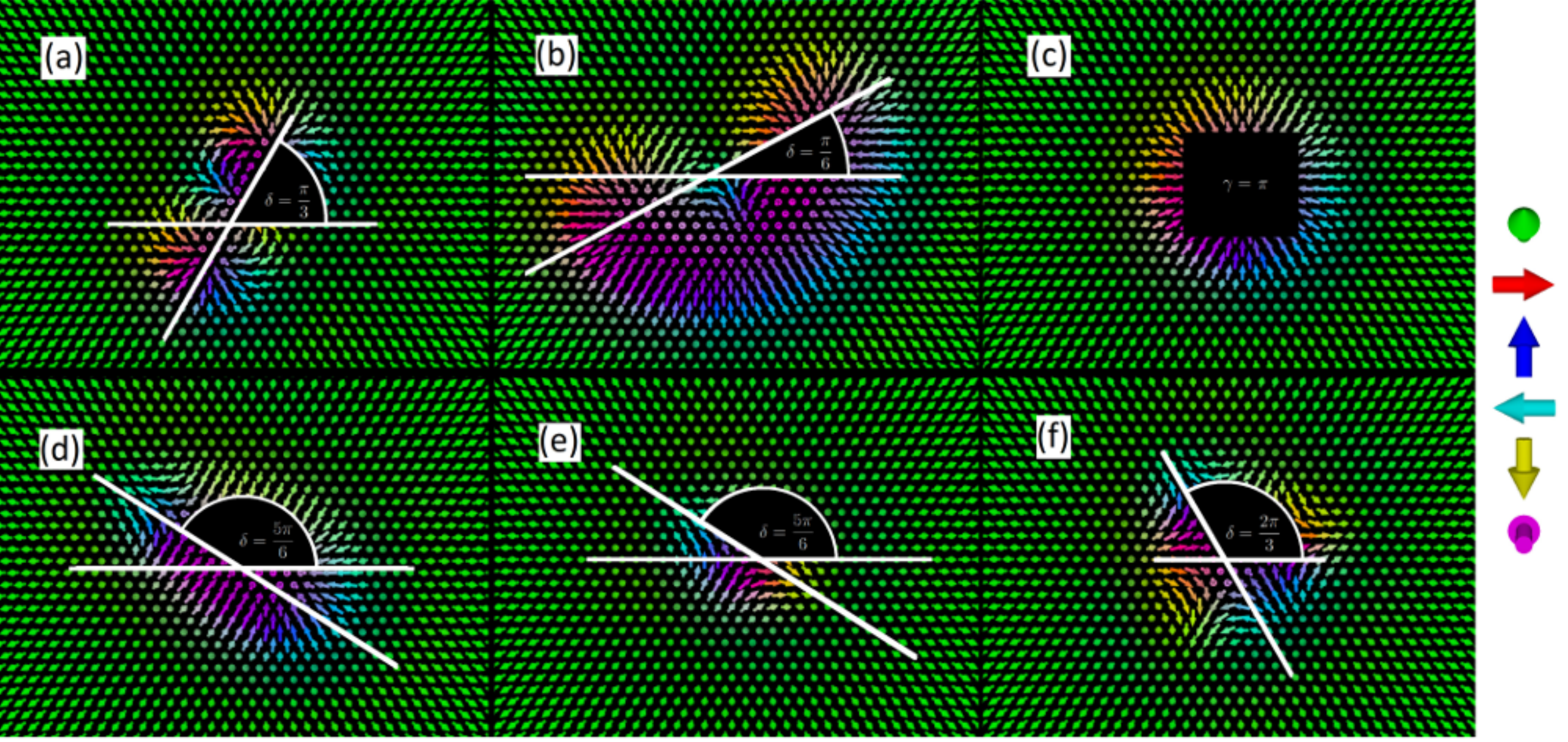}
\caption{Preferred orientation angle $\delta$ of the topological objects in Fig.~\ref{fig3} with respect to the underlying lattice. (a) $\delta=\pi/3$ for $Q=-3$, (b) $\delta=\pi/6$ for $Q=-2$, (d) $\delta=5\pi/6$ for $Q=0$, (e) $\delta=5\pi/6$ for $Q=1$, and (f) $\delta=2\pi/3$ for $Q=2$. Skyrmions with $Q=-1$ in (c) are cylindrically symmetric, and they are characterized by the helicity $\gamma=\pi$ instead.\label{fig3a}}
\end{figure*}

In order to differentiate between the effects caused by the isotropic exchange interactions and the Dzyaloshinsky--Moriya interactions, we have performed the same simulations by replacing the tensorial couplings $\mathcal{J}_{ij}$ in Eq.~(\ref{eqn1}) by only the isotropic Heisenberg couplings $J_{ij}$ in Eq.~(\ref{eqn2}), while modifying the on-site anisotropy tensor $\mathcal{K}$ to keep the total anisotropy energy between the in-plane and out-of-plane orientations the same. In agreement with the theoretical description in Sec.~\ref{sec3} and Ref.~\cite{Lin}, Fig.~\ref{fig6} demonstrates that it is still possible to stabilize all the localized metastable states with finite topological charge in this case, and their shape will correspond to the circular approximation in Eq.~(\ref{eqn9}). However, the ``chimera'' skyrmion has collapsed into the field-polarized state for these interaction parameters.

It can be seen from Table~\ref{table3} that skyrmions and antiskyrmions with opposite topological charges become energetically degenerate, in agreement with the $\left(S^{x},S^{y},S^{z}\right)\rightarrow\left(-S^{x},S^{y},S^{z}\right)$ symmetry of the Hamiltonian (\ref{eqn2}), which switches the sign of the topological charge. Compared to Table~\ref{table2}, it can be seen that the energy of all objects has increased in the absence of the Dzyaloshinsky--Moriya interaction, indicating that skyrmionic structures with $m \neq 1$ also gain energy from the chiral interaction due to their distorted shape. Although higher-order skyrmionic structures possess a higher energy (cf. Ref.~\cite{Lin}), their binding energy is actually negative, meaning that they cannot split into their constituents.

\subsection{Preferred orientation of asymmetric skyrmionic structures on the lattice\label{sec5}}

Besides distorting the shape of skyrmionic structures with $m \neq 1$, the Dzyaloshinsky--Moriya interaction also defines a preferred orientation of these objects with respect to the underlying atomic lattice. We have characterized this orientation by the angle $\delta$ between the $[1\overline{1}0]$ crystallographic direction and a characteristic cross-section of the localized spin configuration, illustrated in Fig.~\ref{fig3a}. Determining the angle $\delta$ is equivalent to defining the helicity $\gamma$ for $m \neq 1$, since the latter also transforms under rotations according to Eq.~(\ref{eqn13}). Furthermore, we note that shifting $\delta$ by $2\pi/3$ leads to an equivalent configuration due to the $C_{3\textrm{v}}$ symmetry of the underlying lattice. For the elongated objects with $Q=-3,-2,0,$ and $1$, we chose the long axis as the characteristic cross-section, yielding the values $\delta=\pi/3,\delta=\pi/6,\delta=5\pi/6,$ and $\delta=5\pi/6$, respectively. For the antiskyrmion with $Q=2$, we chose the symmetry axis of the triangle, for an angle of $\delta\approx 2\pi/3$.

These preferred orientations appear because domain walls along different crystallographic directions possess different energies. As shown in Table~\ref{table4}, domain walls with normal vectors along the $[1\overline{1}0]$ direction are energetically preferred over ones along the $[11\overline{2}]$ in the system; the domain walls are of right-handed N\'{e}el type due to the Dzyaloshinsky--Moriya interaction (DMI). This is in agreement with Ref.~\cite{Rozsa4}, where the same directional preference was found for right-handed cycloidal spin spirals with wave vectors along the different crystallographic axes; the negative domain wall energies indicate that the ground state is actually the spin spiral state. If we consider the model with only the isotropic exchange interactions introduced in Sec.~\ref{sec4a} (no DMI), the ground state becomes ferromagnetic, and the preferred direction for the domain walls switches.

\begin{table}[tbp]
    \centering
\begin{ruledtabular}
    \begin{tabular}{ccc}
normal vector & $\Delta E_{\textrm{DW}} [\textrm{mRy}]$ (DMI) & $\Delta E_{\textrm{DW}} [\textrm{mRy}]$ (no DMI)\\
\hline
$[1\overline{1}0]$ & -0.0211 & 0.2129 \\
$[11\overline{2}]$ & -0.0118 & 0.1167 \\
    \end{tabular}%
\end{ruledtabular}
\caption{Energies of $180^{\circ}$ domain walls along different crystallographic directions for the original Hamiltonian with interaction tensors $\mathcal{J}_{ij}$ containing the Dzyaloshinsky--Moriya interactions (DMI), and by only considering the isotropic exchange interactions $J_{ij}$ (no DMI). The calculations were performed for an $N=128\times128$ lattice with fixed antiparallel boundary conditions along the normal vector of the domain wall, and periodic boundary conditions in the perpendicular direction. The energy differences are normalized to a one-dimensional spin chain.\label{table4}}
\end{table}

As it was discussed in Sec.~\ref{sec3}, topological objects with $m \neq 1$ always possess both right-handed and left-handed segments. Since the Dzyaloshinsky--Moriya interaction switches the preferred domain wall direction for right-handed walls, for left-handed walls the preferred direction must be the same as for the isotropic interactions, since these domain walls lose energy due to the chiral interaction. In this case, the antiskyrmion with $Q=1$ can minimize its energy when its left-handed cross-section is along the $[\overline{1}2\overline{1}]$ axis or a symmetrically equivalent direction (next-nearest neighbors on the lattice), which yields the value $\delta=5\pi/6$ shown in Fig.~\ref{fig3a}(e). Simultaneously, its right-handed cross-section is along the perpendicular $[10\overline{1}]$ direction (nearest neighbors on the lattice), which is also energetically favorable. We could not observe such a preferred orientation when we used only isotropic exchange interactions in the simulations as in Fig.~\ref{fig6}; this is expected as the domain wall energy does not depend on the rotational sense of the spins in this case.

The same argument can be used to explain the orientation of the ``chimera'' skyrmion, since the only difference is that the completely left-handed cross-section is replaced by a pair of left-handed and right-handed $180^{\circ}$ domain walls following each other. In the skyrmion with $Q=-3$, a full $360^{\circ}$ left-handed domain wall can only be observed when moving along its short axis, which is parallel to the $[\overline{1}2\overline{1}]$ direction. Along the long axis, one can observe a $360^{\circ}$ right-handed domain wall, with a shorter segment with reversed chirality in the middle; this is the preferred orientation of right-handed domain walls according to Table~\ref{table4}.

Regarding the skyrmion with $Q=-2$, the above argument would predict that its longer axis, containing mostly right-handed domain walls, would be oriented along the nearest-neighbor direction ($\delta=\pi/3$), while it is parallel to the next-nearest-neighbor direction ($\delta=\pi/6$) in Fig.~\ref{fig3a}(b). This discrepancy may be explained by the very strongly distorted spin configuration, resembling two individual skyrmions along the next-nearest-neighbor direction ($\delta=\pi/6$) which are weakly connected to each other. As it was demonstrated in Ref.~\cite{Rozsa4}, the next-nearest-neighbor direction is preferable for creating bonds between individual skyrmions. Increasing the field to $B=2.35\,\textrm{T}$ (see Table~\ref{table2}) compresses the skyrmion with $Q=-2$ into a more circular shape, and its preferred orientation on the lattice also rotates to $\delta=\pi/3$, which value is in agreement with the prediction based on left-handed and right-handed domain walls.

Finally, we note that the above argument is insufficient for explaining the orientation of the antiskyrmion with $Q=2$ with respect to the lattice, because its cross-section along the symmetry axes of the triangle corresponds to a pair of right-rotating and left-rotating $180^{\circ}$ domain walls.

\section{Conclusion\label{sec6}}

We have examined localized metastable spin configurations in the field-polarized state of (Pt$_{0.95}$Ir$_{0.05}$)/Fe bilayer on Pd$(111)$ surface by using spin dynamics calculations. The interaction parameters in the Hamiltonian have been determined from \textit{ab initio} methods earlier\cite{Rozsa4}. We could identify objects with topological charges $Q=-3,-2,-1,0,1,$ and $2$, and explained their presence by the interplay between the frustrated isotropic exchange interactions and the Dzyaloshinsky--Moriya interaction.

In agreement with the theoretical prediction based on the continuum model, we have demonstrated that the Dzyaloshinsky--Moriya interaction selects skyrmions with $Q=-1$ as the energetically most favorable configuration. However, the other topological objects also remain stable due to the presence of the frustrated isotropic exchange interactions, although their shape becomes distorted because of the chiral interaction, and they assume preferred orientations on the lattice.
We have observed the different skyrmionic structures also for other Ir concentrations $x$ in the (Pt$_{1-x}$Ir$_{x}$)/Fe/Pd(111) system\cite{Rozsa4}, indicating that the stabilization of different topological objects is a consequence of the simultaneous presence of frustrated exchange interactions and the Dzyaloshinsky--Moriya interaction in ultrathin films. The results discussed in this paper may motivate the search for experimental realizations of different topological objects in similar systems.

It was demonstrated in Ref.~\cite{Lin} that if only isotropic exchange interactions are considered, the extra degree of freedom connected to the helicity of the skyrmions significantly influences their current-driven motion. In the system considered in this paper, this continuous symmetry is broken by the presence of the Dzyaloshinsky--Moriya interaction, but rotating the topological objects by $2\pi/3$ still leads to an energetically degenerate configuration due to the symmetry of the lattice. Consequently, this discrete symmetry offers new implications for the current-driven motion of skyrmionic structures with different topological charges.

\begin{acknowledgments}

The authors thank Bertrand Dup\'{e} and Alexei N. Bogdanov for enlightening discussions. Financial support for this work was provided by the Deutsche Forschungsgemeinschaft via SFB 767 ``Controlled Nanosystems: Interaction and Interfacing to the Macroscale'', by the SASPRO Fellowship of the Slovak Academy of Sciences under project no. 1239/02/01, by the Hungarian State E\"{o}tv\"{o}s Fellowship of the Tempus Public Foundation (contract no. 2016-11), and by the National Research, Development and Innovation Office of Hungary under project nos.~K115575 and PD120917.

\end{acknowledgments}


\begin{thebibliography}{1}

	\bibitem{Nagaosa} N. Nagaosa and Y. Tokura, Nature Nanotechnology \textbf{8}, 899 (2013).

	\bibitem{Iwasaki} J. Iwasaki, M. Mochizuki, and N. Nagaosa, Nature Communications \textbf{4}, 1463 (2013).

	\bibitem{Jonietz} F. Jonietz, S. M\"{u}hlbauer, C. Pfleiderer, A. Neubauer, W. M\"{u}nzer, A. Bauer, T. Adams, R. Georgii, P. B\"{o}ni, R. A. Duine, K. Everschor, M. Garst, A. Rosch, Science \textbf{330}, 1648 (2010).

	\bibitem{Fert} A. Fert, V. Cros, and J. Sampaio, Nature Nanotechnology \textbf{8}, 152 (2013).

	\bibitem{Iwasaki2} J. Iwasaki, M. Mochizuki, and N. Nagaosa, Nature Nanotechnology \textbf{8}, 742 (2013).

	\bibitem{Zhou} Y. Zhou and M. Ezawa, Nature Communications \textbf{5}, 4652 (2014).

	\bibitem{Jiang} W. Jiang, P. Upadhyaya, W. Zhang, G. Yu, M. B. Jungfleisch, F. Y. Fradin, J. E. Pearson, Y. Tserkovnyak, K. L. Wang, O. Heinonen, S. G. E. te Velthuis, and A. Hoffmann, Science \textbf{349}, 283 (2015).

	\bibitem{Woo} S. Woo, K. Litzius, B. Kr\"{u}ger, M.-Y. Im, L. Caretta, K. Richter, M. Mann, A. Krone, R. M. Reeve, M. Weigand, P. Agrawal, I. Lemesh, M.-A. Mawass, P. Fischer, M. Kl\"{a}ui, and G. S. D. Beach, Nature Materials \textbf{15}, 501 (2016).

	\bibitem{Belavin} A. A. Belavin and A. M. Polyakov, Pis'ma Zh. Eksp. Teor. Fiz. \textbf{22}, 503 (1975) [Sov. Phys. JETP Lett. \textbf{22}, 245 (1975)].

	\bibitem{Dzyaloshinsky} I. Dzyaloshinsky, J. Phys. Chem. Solids \textbf{4}, 241 (1958).

	\bibitem{Moriya} T. Moriya, Phys. Rev. Lett. \textbf{4}, 228 (1960).

	\bibitem{Bogdanov2} A. Bogdanov and A. Hubert, J. Magn. Magn. Mater. \textbf{138}, 255 (1994).

	\bibitem{Okubo} T. Okubo, S. Chung, and H. Kawamura, Phys. Rev. Lett. \textbf{108}, 017206 (2012).

	\bibitem{Heinze} S. Heinze, K. von Bergmann, M. Menzel, J. Brede, A. Kubetzka, R. Wiesendanger, G. Bihlmayer, and S. Bl\"{u}gel, Nature Physics \textbf{7}, 713 (2011).

	\bibitem{Malozemoff} A. P. Malozemoff and J. C. Slonczewski, \textit{Magnetic Domain Walls in Bubble Materials} (Academic Press, New York, 1979).

	\bibitem{Yu3} X. Yu, M. Mostovoy, Y. Tokunaga, W. Zhang, K. Kimoto, Y. Matsui, Y. Kaneko, N. Nagaosa, and Y. Tokura, Proc. Natl. Acad. Sci. USA \textbf{109}, 8856 (2012).

	\bibitem{Yu4} X. Z. Yu, Y. Tokunaga, Y. Kaneko, W. Z. Zhang, K. Kimoto, Y. Matsui, Y. Taguchi, and Y. Tokura, Nature Communications \textbf{5}, 3198 (2014).

	\bibitem{Kiselev} N. S. Kiselev, A. N. Bogdanov, R. Sch\"{a}fer, and U. K. R\"{o}ssler, J. Phys. D: Appl. Phys. \textbf{44}, 392001 (2011).

	\bibitem{Muhlbauer} S. M\"{u}hlbauer, B. Binz, F. Jonietz, C. Pfleiderer, A. Rosch, A. Neubauer, R. Georgii, and P. B\"{o}ni, Science \textbf{323}, 915 (2009).

	\bibitem{Yu2} X. Z. Yu, N. Kanazawa, Y. Onose, K. Kimoto, W. Z. Zhang, S. Ishiwata, Y. Matsui, and Y. Tokura, Nature Materials \textbf{10}, 106 (2010).

	\bibitem{Wilhelm} H. Wilhelm, M. Baenitz, M. Schmidt, U. K. R\"{o}ssler, A. A. Leonov, and A. N. Bogdanov, Phys. Rev. Lett. \textbf{107}, 127203 (2011).

	\bibitem{Munzer} W. M\"{u}nzer, A. Neubauer, T. Adams, S. M\"{u}hlbauer, C. Franz, F. Jonietz, R. Georgii, P. B\"{o}ni, B. Pedersen, M. Schmidt, A. Rosch, and C. Pfleiderer, Phys. Rev. B \textbf{81}, 041203(R) (2010).

	\bibitem{Yu} X. Z. Yu, Y. Onose, N. Kanazawa, J. H. Park, J. H. Han, Y. Matsui, N. Nagaosa, and Y. Tokura, Nature \textbf{465}, 901 (2010).

	\bibitem{Adams} T. Adams, A. Chacon, M. Wagner, A. Bauer, G. Brandl, B. Pedersen, H. Berger, P. Lemmens, and C. Pfleiderer, Phys. Rev. Lett. \textbf{108}, 237204 (2012).

	\bibitem{Kezsmarki} I. K\'{e}zsm\'{a}rki, S. Bord\'{a}cs, P. Milde, E. Neuber, L. M. Eng, J. S. White, H. M. R{\o}nnow, C. D. Dewhurst, M. Mochizuki, K. Yanai, H. Nakamura, D. Ehlers, V. Tsurkan, and A. Loidl, Nature Materials \textbf{14}, 1116 (2015).

	\bibitem{Tokunaga} Y. Tokunaga, X. Z. Yu, J. S. White, H. M. R{\o}nnow, D. Morikawa, Y. Taguchi, and Y. Tokura, Nature Communications \textbf{6}, 7638 (2015).

	\bibitem{Bogdanov} A. N. Bogdanov and D. A. Yablonski\u{i}, Zh. Eksp. Teor. Fiz. \textbf{95}, 178 (1989) [Sov. Phys. JETP \textbf{68}, 101 (1989)].

	\bibitem{Dupe} B. Dup\'{e}, M. Hoffmann, Ch. Paillard, and S. Heinze, Nature Communications \textbf{5}, 4030 (2014).

	\bibitem{Simon} E. Simon, K. Palot\'{a}s, L. R\'{o}zsa, L. Udvardi, and L. Szunyogh, Phys. Rev. B \textbf{90}, 094410 (2014).

	\bibitem{Leonov2} A. O. Leonov, T. L. Monchesky, N. Romming, A. Kubetzka, A. N. Bogdanov, and R. Wiesendanger, New J. Phys. \textbf{18}, 065003 (2016).

	\bibitem{Romming} N. Romming, C. Hanneken, M. Menzel, J. E. Bickel, B. Wolter, K. von Bergmann, A. Kubetzka, and R. Wiesendanger, Science \textbf{341}, 636 (2013).

	\bibitem{Hsu} P.-J. Hsu, A. Kubetzka, A. Finco, N. Romming, K. von Bergmann, and R. Wiesendanger, Nature Nanotechnology (2016), doi:10.1038/nnano.2016.234.

	\bibitem{Moreau-Luchaire} C. Moreau-Luchaire, C. Moutafis, N. Reyren, J. Sampaio, C. A. F. Vaz, N. Van Horne, K. Bouzehouane, K. Garcia, C. Deranlot, P. Warnicke, P. Wohlh\"{u}ter, J.-M. George, M. Weigand, J. Raabe, V. Cros, A. Fert, Nature Nanotechnology \textbf{11}, 444 (2016).

	\bibitem{Dupe2} B. Dup\'{e}, G. Bihlmayer, M. B\"{o}ttcher, S. Bl\"{u}gel, and S. Heinze, Nature Communications \textbf{7}, 11779 (2016).

	\bibitem{Yang} H. Yang, A. Thiaville, S. Rohart, A. Fert, and M. Chshiev, Phys. Rev. Lett. \textbf{115}, 267210 (2015).

	\bibitem{Leonov} A. O. Leonov and M. Mostovoy, Nature Communications \textbf{6}, 8275 (2015).

	\bibitem{Lin} S.-Z. Lin and S. Hayami, Phys. Rev. B \textbf{93}, 064430 (2016).

	\bibitem{Lee} J. C. T. Lee, J. J. Chess, S. A. Montoya, X. Shi, N. Tamura, S. K. Mishra, P. Fischer, B. J. McMorran, S. K. Sinha, E. E. Fullerton, S. D. Kevan, and S. Roy, Appl. Phys. Lett. \textbf{109}, 022402 (2016).

	\bibitem{Wang} W. Wang, Y. Zhang, G. Xu, L. Peng, B. Ding, Y. Wang, Z. Hou, X. Zhang, X. Li, E. Liu, S. Wang, J. Cai, F. Wang, J. Li, F. Hu, G. Wu, B. Shen, and X.-X. Zhang, Adv. Mat. \textbf{28}, 6887 (2016).

	\bibitem{Dupe3} B. Dup\'{e}, C. N. Kruse, T. Dornheim, and S. Heinze, New J. Phys. \textbf{18}, 055015 (2016).

	\bibitem{Rozsa3} L. R\'{o}zsa, L. Udvardi, L. Szunyogh, and I. A. Szab\'{o}, Phys. Rev. B \textbf{91}, 144424 (2015).

	\bibitem{Romming3} N. Romming, M. Hoffmann, B. Dup\'{e}, K. von Bergmann, S. von Malottki, A. Kubetzka, R. Wiesendanger, and S. Heinze, arXiv:1610.07853 (2016).

	\bibitem{Rozsa4} L. R\'{o}zsa, A. De\'{a}k, E. Simon, R. Yanes, L. Udvardi, L. Szunyogh, and U. Nowak, Phys. Rev. Lett. \textbf{117}, 157205 (2016).


	\bibitem{Szunyogh} L. Szunyogh, B. \'{U}jfalussy, P. Weinberger, and J. Koll\'{a}r, Phys. Rev. B \textbf{49}, 2721 (1994).

	\bibitem{Zeller} R. Zeller, P. H. Dederichs, B. \'{U}jfalussy, L. Szunyogh, and P. Weinberger, Phys. Rev. B \textbf{52}, 8807 (1995).

	\bibitem{Udvardi} L. Udvardi, L. Szunyogh, K. Palot\'{a}s, and P. Weinberger, Phys. Rev. B \textbf{68}, 104436 (2003).

	\bibitem{Nowak} U. Nowak, in \textit{Handbook of Magnetism and Advanced Magnetic Materials}, edited by H. Kronm\"{u}ller and S. Parkin, Vol. 2 (Wiley, New York, 2007).

	\bibitem{Mentink} J. H. Mentink, M. V. Tretyakov, A. Fasolino, M. I. Katsnelson, and Th. Rasing, J. Phys.: Condens. Matter \textbf{22}, 176001 (2010).

	\bibitem{Michelson} A. Michelson, Phys. Rev. B \textbf{16}, 577 (1977).

	\bibitem{Rozsa2} L. R\'{o}zsa, E. Simon, K. Palot\'{a}s, L. Udvardi, and L. Szunyogh, Phys. Rev. B \textbf{93}, 024417 (2016).

	\bibitem{Berg} B. Berg and M. L\"{u}scher, Nucl. Phys. B \textbf{190}, 412 (1981).

	\bibitem{Hagemeister} J. Hagemeister, N. Romming, K. von Bergmann, E. Y. Vedmedenko, and R. Wiesendanger, Nature Communications \textbf{6}, 8455 (2015).

	\bibitem{Lobanov} I. S. Lobanov, H. J\'{o}nsson, and V. M. Uzdin, Phys. Rev. B \textbf{94}, 174418 (2016).

\end{thebibliography}
\end{document}